\documentclass[preprint,showpacs,preprintnumbers,amsmath,pre]{revtex4}
\usepackage{graphicx}
\newcommand{\y }{\'{\i}}
\newcommand{\be }{\begin{equation}}
\newcommand{\ee }{\end{equation}}

\begin{document}

\title{Response of a catalytic reaction to periodic variation of the
CO pressure: Increased CO$_2$ production and dynamic phase transition}
\author{Erik Machado}
\author{Gloria M.~Buend\y a}
\affiliation{Physics Department, Universidad Sim\'on Bol\y var,\\
Apartado 89000, Caracas 1080, Venezuela}
\author{Per Arne Rikvold}
\affiliation{
Center for Materials Research and Technology,
School of Computational Science, 
and Department of Physics, \\
Florida State University, Tallahassee, Florida 32306-4052, USA}
\author{Robert M.\ Ziff}
\affiliation{
Department of Chemical Engineering and Michigan Center for Theoretical 
Physics, University of Michigan, Ann Arbor, MI 48109-2136, USA
}

\date{\today}
\begin{abstract}
We present a kinetic Monte Carlo study of the dynamical response of a
Ziff-Gulari-Barshad model for CO oxidation with CO desorption to periodic
variation of the CO pressure. We use a square-wave periodic
pressure variation with parameters that can be tuned to enhance the
catalytic activity. We produce evidence that, below
a critical value of the desorption rate, the driven system undergoes a dynamic
phase transition between a CO$_2$ productive phase and a
nonproductive one at a critical value of the period and waveform of the pressure 
oscillation. At the dynamic phase transition the period-averaged 
CO$_2$ production rate is 
significantly increased and can be used as a dynamic order parameter. 
We perform a finite-size scaling analysis that
indicates the existence of power-law singularities for the order parameter 
and its fluctuations, yielding estimated critical exponent ratios 
$\beta/\nu \approx 0.12$ and $\gamma/\nu \approx 1.77$. 
These exponent ratios, together with theoretical symmetry arguments and
numerical data for the fourth-order cumulant associated with the transition, 
give reasonable support for 
the hypothesis that the observed nonequilibrium dynamic 
phase transition is in the same universality class as the two-dimensional 
equilibrium Ising model. 
\end{abstract}

\pacs{
64.60.Ht, 
82.65.+r, 
82.20.Wt, 
05.40.-a 
}

\maketitle

\section{Introduction}

The study of nonequilibrium statistical models is a subject of
current interest in a broad range of fields such as chemical
reactions, fluid turbulence, chaos, biological populations,
growth-deposition processes, and even
economics \cite{general1,general2}. In particular, the study of
surface reaction systems has received a great deal of attention
\cite{surface}. These systems not only constitute a fruitful
laboratory for exploring critical phenomena associated with
out-of-equilibrium statistical physics, but they can also play an
important role in the development of more efficient catalytic
processes. Catalytic reactions are widespread in nature and have
many industrial and technological applications \cite{catalytic,
imbhil95}.

Several experiments show that it is possible to increase the
efficiency of catalytic reactions by subjecting the system to
periodic external forcing
\cite{imbhil95,ehsasi89,vaporciyan88,cutlip83,hegedus80}.
Monte Carlo simulations of the
Ziff, Gulari, and Barshad (ZGB) model without desorption
\cite{ziff86} indicate that an enhancement of the catalytic
activity is observed when the system is perturbed by a periodic
force that drives it briefly into the CO poisoned
state \cite{ziff86,lopez00}. However, it is well known that the ZGB model
does not reproduce several important aspects of catalytic
processes, such as lateral diffusion \cite{TAMM98} 
and CO desorption \cite{tome93}. 
In the present study we neglect diffusion and concentrate on the 
effects of CO desorption. For brevity we will refer to this model
as the ZGB-k model \cite{tome93}. 

Other work by the present authors \cite{machado04} indicates that near the
coexistence line between the active and the CO poisoned regime,
the decay times of the metastable states are different if the
ZGB-k model is driven into the CO poisoned state from the active
phase, or if it is driven into the active phase from the CO
poisoned state. Based on this result, we expect that the catalytic
activity of the system will increase when the system is subjected
to periodic variation of the external CO pressure, only when one takes into
account that the time it takes the system to decontaminate is
different from the time it takes it to contaminate. 
Furthermore, we show that the ZGB-k model, driven by 
an oscillating CO pressure, undergoes a dynamic phase transition, similar 
to the one observed in Ising 
\cite{TOME90,MEND91,CHAK99,ising1,ising2,FUJI01,KORN02}, 
anisotropic Heisenberg 
\cite{JANG01,JANG03},
and $XY$ models 
\cite{YASU02}, 
driven by an oscillating applied field. 

The rest of this paper is organized as follows. In Sec.~II we
define the model and describe the Monte Carlo simulation
techniques used. In Sec.~III we present and discuss the numerical results
obtained when subjecting the model to a periodic variation of the external CO
pressure. Finally, we present our conclusions in Sec.~IV.

\section{Model and simulations}

The original ZGB model without desorption \cite{ziff86} describes
some kinetic aspects of the reaction CO+O $\rightarrow$ CO$_2$ on
a catalytic surface in terms of a single parameter: the probability $y$ 
that the next molecule arriving at the surface is CO. This parameter is 
proportional to the partial pressure  of CO and will loosely be referred to 
as the CO pressure. The model exhibits two
transitions, a continuous one at low CO pressure to an
oxygen-saturated surface, and a discontinuous one at a higher CO
pressure to a CO saturated surface. When desorption is included,
the distinction between the high and low CO coverage phases
disappears for a desorption rate $k$ above a critical value, 
$k_c \approx 0.0406$ \cite{tome93,machado04}. 
There is evidence that the nonequilibrium phase transition at $k_c$ 
is in the same universality class as the two-dimensional kinetic Ising 
model at equilibrium \cite{tome93}. 
Below $k_c$ there are well-defined low and high
coverage phases, separated by a discontinuous, 
nonequilibrium phase transition. 
This picture is consistent with experimental
observations on catalytic surfaces that show transitions between
low and high coverage phases ~\cite{ehsasi89, matsushima79, golchet78, christmann73}.

The ZGB model with CO desorption is simulated on a square lattice
of side $L$ that represents the catalytic surface. A Monte Carlo
simulation generates a sequence of trials: adsorption with
probability $1-k$ and desorption with probability $k$. A site is
selected at random. In desorption, if the site is occupied by
a CO it is vacated, if not the trial ends. In adsorption, if
the site is occupied the trial ends, if not a CO or O$_2$ molecule
is selected with probability $y$ or $1-y$ (the relative
impingement rates of CO and O$_2$), respectively. 
A CO molecule can be adsorbed at the empty
site if none of its nearest neighbors are occupied by an O atom.
Otherwise, one of the occupied neighbors is selected at random and
removed from the surface, liberating a CO$_2$ molecule. O$_2$
molecules require a nearest-neighbor pair of vacant sites to
adsorb. Once an O$_2$ molecule is adsorbed, it dissociates into
two O atoms. If an O atom is located next to a site filled with a
CO molecule, they react to form a CO$_2$ molecule that escapes,
leaving two sites vacant. This process mimics the CO + O $\rightarrow$ CO$_2$
surface reaction.

\section{Results}

Our simulations were performed on a square lattice of
$L{\times}L$ sites, assuming periodic boundary conditions. The
time unit is one Monte Carlo step per site (MCSS), in which each
site is visited once, on average.

\begin{figure}
\centering\includegraphics[scale=0.44]{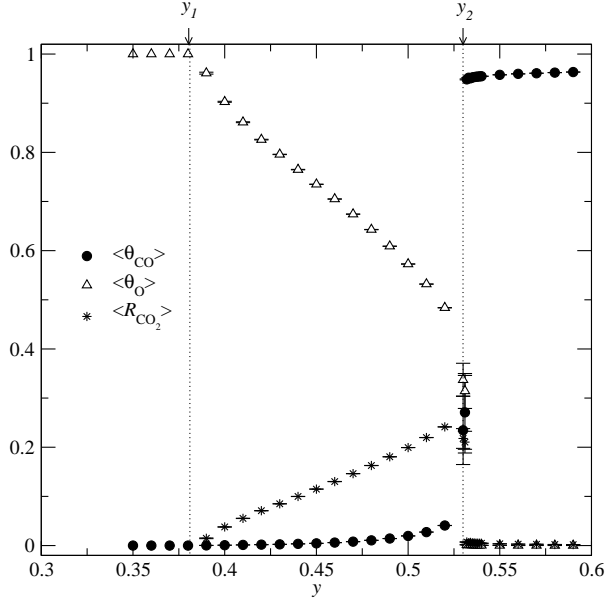}
\caption{Average values of the CO and  O coverages, $\theta_{\rm CO}$ 
and $\theta_{\rm O}$, and of the
CO$_2$ production rate, $R_{\rm CO_2}$, 
shown as functions of the stationary applied CO pressure
$y$, for $L=100$ with $k=0.02$. A continuous, nonequilibrium 
phase transition occurs at $y_1$, and a discontinuous one at $y_2(k)$.}
\label{stat1b}
\end{figure}

The coverages $\theta_\text{CO}$ and $\theta_\text{O}$ are defined
as the fraction of surface sites occupied by CO and O,
respectively, and $R_{\text{CO}_2}$ is defined as the rate of
production of CO$_{2}$. In Fig.~\ref{stat1b} we show the
dependence on the external constant CO pressure $y$
of the average value of the coverages and the
CO$_2$ production rate: $\langle \theta_\text{CO} \rangle$,
$\langle \theta_\text{O} \rangle$ and $\langle R_{\text{CO}_2} \rangle$,
respectively. There are two inactive
regions, $y<y_1$ ($y_1$ seems to be fairly independent of $k$, as
expected because of the absence of CO on the surface at this point) and
$y>y_2(k)$, corresponding to the cases in which the surface is
saturated with O and CO, respectively. For
$y_{1}<y<y_{2}(k)$, there is an active window where the system
produces CO$_{2}$. Notice that the maximum value of
$\langle R_{\text{CO}_2} \rangle$ is reached as $y_2(k)$
is approached from the low-$\theta_\text{CO}$ phase.

One way to perturb the catalytic oxidation process in order to
increase the CO$_2$ production is by switching the external
pressure back and forth across the discontinuous transition
$y_2(k)$ \cite{imbhil95,ehsasi89,vaporciyan88,cutlip83,hegedus80}.
It has been shown that when the ZGB system without desorption is
quickly oscillated between the low and high $\theta_\text{CO}$ 
phases by a periodic variation of the external pressure, there is 
considerable enhancement of the catalytic activity \cite{ziff86,lopez00}.
The period and amplitude of the driving pressure must be
calibrated very carefully in order to avoid driving the system
irreversibly toward the CO poisoned state.

\begin{figure}
\centering\includegraphics[scale=0.45]{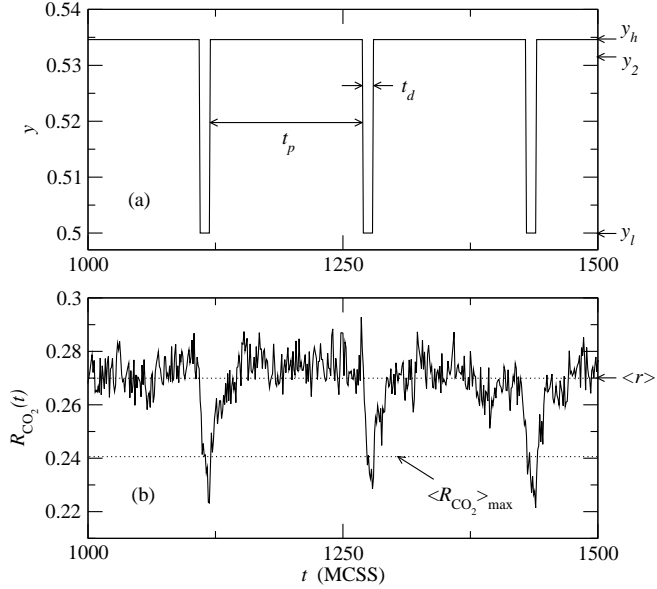}
\caption{(a) Applied periodic pressure of CO, $y(t)$, 
that takes the values
$y_l=0.5$ and $y_h=0.535$ during the time intervals $t_d=10$ and
$t_p=150$, respectively.
(b) Response of the production rate to the
applied pressure given in (a) for $L=100$ with $k=0.02$.
The dotted line marked $\langle r \rangle $ 
indicates the long-time average of the 
period-averaged CO$_2$ production
rate $r$, defined by Eq.~(\protect\ref{order_parameter}), 
while the dotted line marked $\langle R_{\rm CO_2} \rangle_{\rm max}$ marks 
the maximum average CO$_2$ production rate for {\it constant\/} $y=y_2(k)$, 
shown in Fig.~\protect\ref{stat1b}.
Time is measured in units of MCSS.}
\label{serie_osc}
\end{figure}

In other work \cite{machado04} we have calculated the lifetimes
associated with the decay of the metastable states of the ZGB-k
model. The system was prepared in the low (high) CO coverage phase
with an initial pressure $y_i<y_2(k)$ ($y_i>y_2(k))$, and then $y$
was suddenly changed to $y_f>y_2(k)$ ($y_f<y_2(k)$). We then
measured the time it took the system to leave the metastable
state in both cases. We found that the lifetimes depend on the
direction of the process, the decontamination time $\tau_d$
(from high to low CO coverage) being different from the poisoning time
$\tau_p$ (from low to high CO coverage). Inspired by this result,
we decided to subject the system to an oscillating pressure $y(t)$
that in a period $T=t_{d}+t_{p}$ takes the values ,
\begin{equation}
y=\left\{
\begin{array}{ll}
y_l & \mbox{during the time interval}\ t_d \\
y_h & \mbox{during the time interval}\ t_p
\;,
\end{array}
\right.
\label{y_vs_t}
\end{equation}
located at both sides of the transition point, i.e., $y_l < y_2(k)<y_h$
(see Fig.~\ref{serie_osc}(a)).
We found that, for each selection of $y_l$ and $y_h$, by tuning
$t_{d}$ and $t_{p}$, the times that the driving force spends in
the low and high coverage regions respectively, we could increase
the productivity of the system.  In Fig~\ref{serie_osc}(b) it is
seen that the response to the periodic pressure
shown in Fig~\ref{serie_osc}(a), the CO$_2$ production rate $R_{\text{CO}_2}$, 
also exhibits an oscillatory behavior.
We therefore use its period-averaged value, defined as
\begin{equation}
r = \frac{1}{T}\oint R_{\text{CO}_2}(t)dt \;,
\label{order_parameter}
\end{equation}
as the dynamic order parameter. For the parameters used in 
Fig.~\ref{serie_osc}, the long-time average of $r$ is 
$\langle r \rangle = 0.2683$, 11\% higher than 
the maximum average CO$_2$ production rate for constant $y$, 
$\langle R_{\rm CO_2} \rangle_{\rm max}=0.2414$. 
(Compare Fig.~\ref{stat1b} and Fig~\ref{serie_osc}(b).)

\begin{figure}
\centering\includegraphics[scale=0.5]{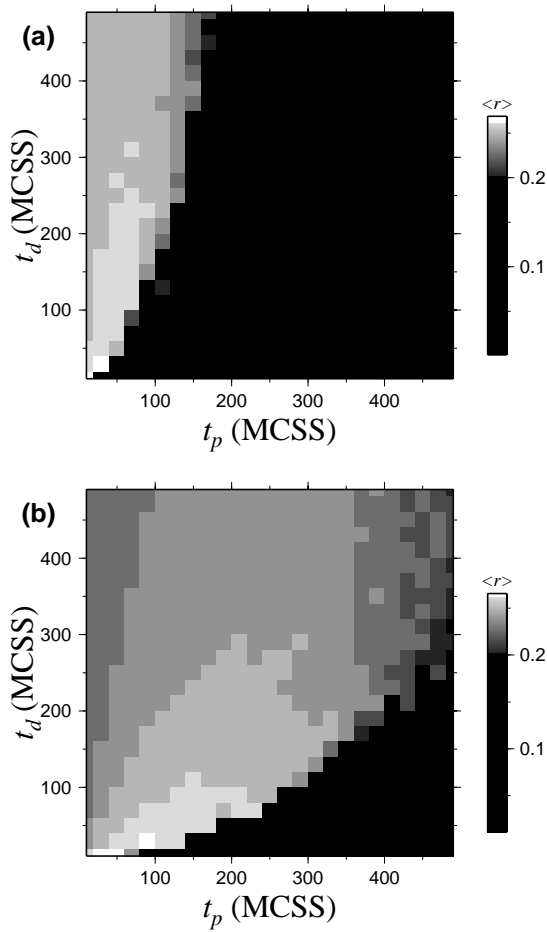}
\caption{
Long-time average $\langle r\rangle$ of the period-averaged 
CO$_2$ production rate $r$, shown as a density plot vs 
$t_d$ and $t_p$ for (a) $y_l=0.52$, $y_h=0.54$,
and (b) $y_l=0.51$, $y_h=0.535$. $k=0.01$ and $L=100$.}
\label{imgs}
\end{figure}

By averaging $r$ over many periods of oscillation 
(of the order of $2\times 10^3$),
we found that, depending on the values of $t_d$ and $t_p$, the
system has two well-defined regimes: a productive one with 
$\langle r\rangle >
0$, and a nonproductive one with $\langle r\rangle\approx 0$, 
separated by a transition line in the ($t_p,t_d$) plane. 
Fig.~\ref{imgs} shows density plots of $\langle r\rangle $ in terms of $t_d$ and $t_p$
for two choices of $y_l$ and $y_h$.

\begin{figure}
\centering\includegraphics[scale=0.44]{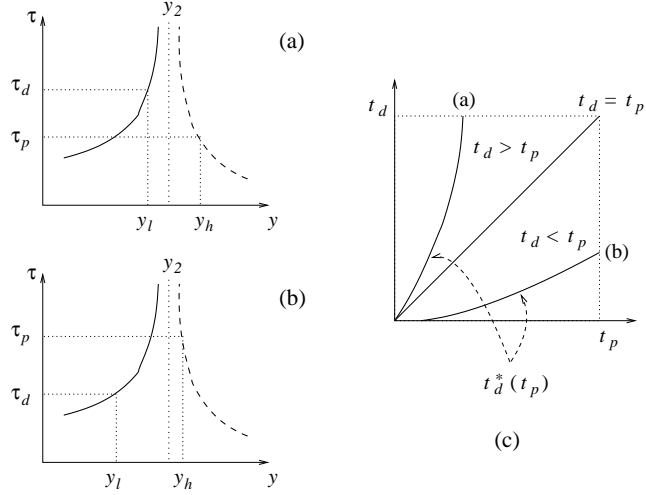} \caption{(a) and (b):
Schematic representations of the decay times of the metastable
states as functions of the CO pressure $y$. The asymmetry of the
curves indicates that the lifetimes depend on the direction of
approach to the coexistence point, $y_2(k)$. The ratio
$\tau_d/\tau_p$ depends on the values of $y_l$ and $y_h$, which are 
different in (a) and (b). 
(a) corresponds to the situation shown in Fig.~\ref{imgs} (a), and 
(b) corresponds to Fig.~\ref{imgs} (b). 
(c) Schematic representation of the dependence of the transition line
between the phases with $\langle r \rangle > 0$ and 
$\langle r \rangle\approx 0$, $t_d^\ast (t_p)$, on
how $y_l$ and $y_h$ are selected.} 
\label{meta}
\end{figure}

The transition line depends strongly on the selected values of
$y_l$ and $y_h$, as can be seen by comparing Fig.~\ref{imgs} (a) and
Fig.~\ref{imgs} (b). This behavior can be understood by looking at
the dependence of the lifetimes on the pressure. In
Fig.~\ref{meta} we show a schematic, but qualitatively correct,
diagram of the lifetimes $\tau_p$ and $\tau_d$ vs $y$. The
lifetimes increase rapidly as $y$ approaches the coexistence point
$y_2(k)$, but the dependences of $\tau_d$ and $\tau_p$ on the
distance to $y_2(k)$ are not equal (this is evident in the
limiting case, $k\rightarrow 0$, where $\tau_d\rightarrow \infty$
independent of $y$), as the lifetimes depend on the direction of
the decay. For the values of $y_l$ and $y_h$ selected in
Fig.~\ref{meta}(a), the system takes longer to be decontaminated
than to be poisoned, i.e., $\tau_d>\tau_p$,
suggesting that the transition line (corresponding to the 
region of high CO$_2$ production) lies in the region $t_d>t_p$, 
as can be seen in Fig.~\ref{imgs} (a). When $y_l$ and $y_h$ are
selected as in Fig.~\ref{meta}(b), the system takes less time to be 
decontaminated than to be poisoned, i.e., $\tau_d<\tau_p$. Then the
transition line is expected to lie in the region $t_d<t_p$, as
seen in Fig.~\ref{imgs} (b). The location of the transition line 
in each case is schematically indicated in Fig.~\ref{meta}(c).

\begin{figure}
\centering\includegraphics[clip, scale=0.45]{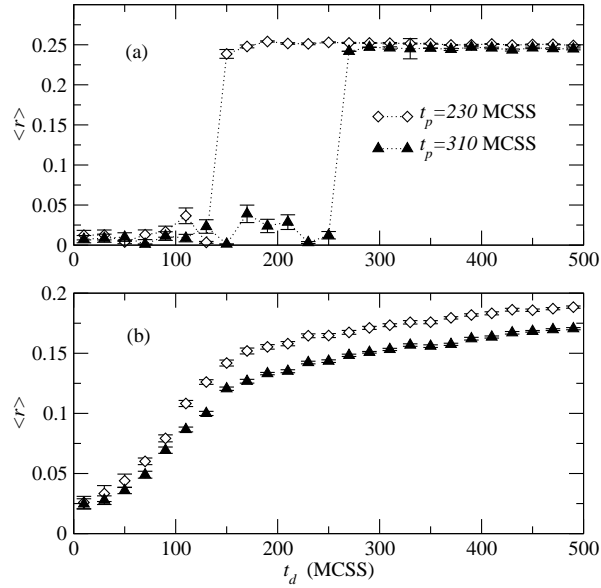}
\caption{Long-time average of the period-averaged rate of CO$_2$ production, 
$\langle r\rangle$, shown vs $t_d$ for two values of $t_p$ and $L=100$;
(a) with $y_l=0.52$, $y_h=0.535$, 
and $k=0.01$, and (b) with $y_l=0.52$, $y_h=0.553$, and $k=0.04$.
Only for $k=0.01$ the system clearly presents two dynamic phases: 
one with $\langle r\rangle\approx 0$ and
the other with $\langle r\rangle>0$.}
\label{r_vs_td}
\end{figure}

\begin{figure}
\centering\includegraphics[scale=0.48]{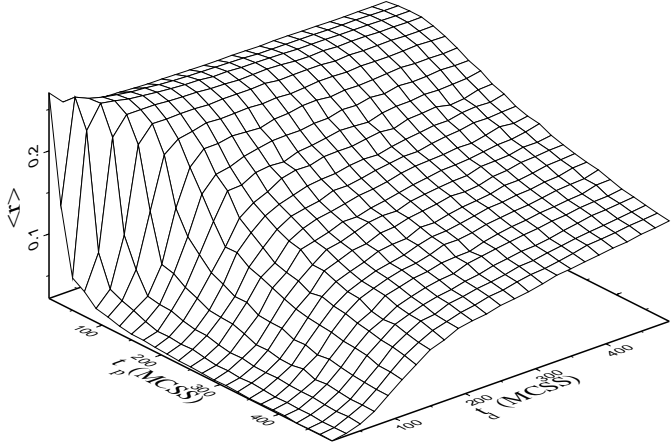}
\caption{Surface plot of $\langle r\rangle$ vs $t_d$ and
$t_p$ for $y_l=0.52$, $y_h=0.553$, and $k=0.04$. $L=100$.}
\label{srf432}
\end{figure}

In Fig.~\ref{r_vs_td} it can be seen how
the average CO$_2$ production, $\langle r\rangle$, depends on $t_d$ for two
different values of $t_p$ and $k$. When $k=0.01$,
Fig.~\ref{r_vs_td} (a) shows [consistent with Fig.~\ref{imgs} (a) and
Fig.~\ref{imgs} (b)] that the system has two well-defined
dynamic phases, one with $\langle r\rangle > 0$ and the other with 
$\langle r\rangle\approx 0$. When
$k$ is increased to 0.04, the system changes continuously from one
phase to the other, as seen in Fig.~\ref{r_vs_td} (b) and
Fig.~\ref{srf432}. This behavior suggests that for low values of
$k$, the system has a dynamic phase transition (DPT)
between a high CO$_2$ productive dynamic phase and a nonproductive one.
This is reminiscent of the behavior of ferromagnetic 
Ising or anisotropic $XY$ or Heisenberg spin
systems driven by a periodically oscillating field 
\cite{TOME90,MEND91,ACHAR97_1,ACHAR97_2,CHAK99,
ising1,ising2,FUJI01,KORN02,JANG01,JANG03,YASU02}.

Universality and finite-size scaling are well-known tools to
analyze critical phenomena. Near a second-order 
equilibrium phase transition, the order-parameter correlation length 
diverges, leading to power-law singularities in terms of 
the finite system size $L$ \cite{pheno}. 
Following the lead of previous applications to the DPT in the kinetic Ising 
model driven by an oscillating field \cite{ising1,ising2}, we use
the period-averaged CO$_2$ production rate $r$ 
as a dynamic order parameter and perform a finite-size scaling analysis of 
its fluctuations and mean value. We define a measure of the
fluctuations in $r$ in a $L \times L$ system in the standard way as
\begin{equation}
X_L=L^{2}[ \langle r^2\rangle - \langle r \rangle^2]
\end{equation}
and measure $X_L$ as a function of $t_d$ for a
fixed value of $t_p$ and several values of $L$. Analogous to the 
situation at a second-order equilibrium phase 
transition, the order-parameter fluctuations increase with
the system size, such that the maximum value of $X_L$ scales as
$X_L^\text{max}\sim L^{\gamma/\nu}$, and the $n$th moment of the
order parameter at the transition scales as
$\langle r^n\rangle_L \sim L^{- n(\beta/\nu)}$.
(We use standard notation for the critical exponents: 
$\gamma$ for the fluctuation exponent, $\beta$ for the order-parameter 
exponent, and $\nu$ for the correlation-length exponent \cite{GOLD92}.)

\begin{figure}
\centering\includegraphics[clip, scale=0.44]{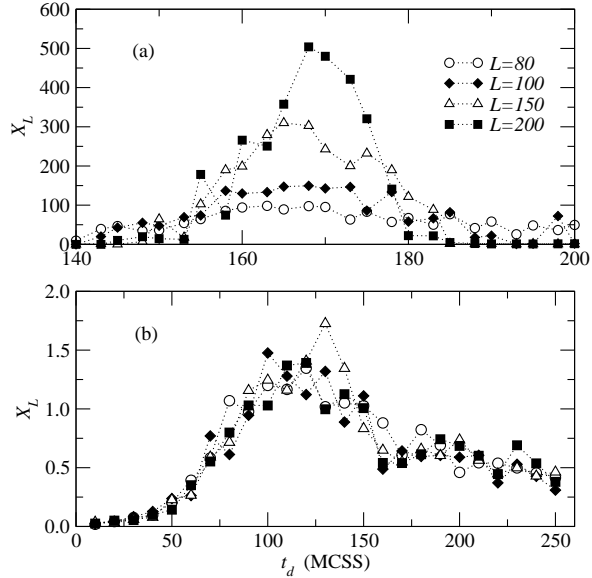}
\caption{The order-parameter fluctuation measure
$X_L$, shown vs $t_d$ for $t_p=230$, for four system
sizes, (a) $k=0.01$ and (b) $k=0.04 \approx k_c$.
The dotted lines are guides to the eye. For clarity the error bars are not
included in the plot. The highest values of $X_L$ have an error of approximately 6 \%.}
\label{var_r_vs_td}
\end{figure}

In Fig.~\ref{var_r_vs_td} we show $X_L$ vs $t_d$ for four system
sizes at $k=0.01$ and $k=0.04$. The errors in $X_L$  are obtained by
standard error propagation analysis as $\sigma_{X_L}\approx 2X_L/\sqrt{n-1}$.
Fig.~\ref{var_r_vs_td} (a) shows that, for $k=0.01$ and the four
values of $L$ used, $X_L$ displays a clear peak, which increases in
height with increasing $L$. This is in clear contrast with
Fig.~\ref{var_r_vs_td} (b) for $k=0.04 \approx k_c$, which shows no such
increase with $L$.

\begin{figure}
\centering\includegraphics[clip, scale=0.44]{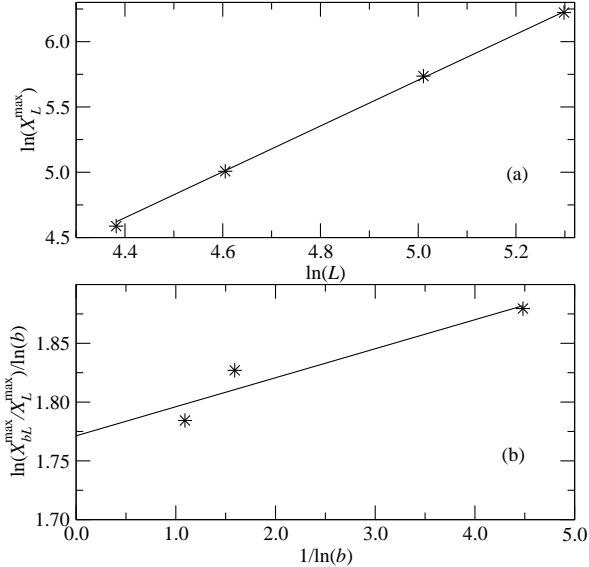}
\caption{(a) Plot of $\ln(X_L^\text{max})$ vs $\ln(L)$ and
(b) plot of $\ln(X_{bL}^\text{max}/X_L^\text{max})/\ln(b)$ vs $1/\ln(b)$,  
both for $k=0.01$ and
$t_p=230$, and including all four system sizes. 
$X_L^\text{max}$ is the maximum value
of $X_L$, taken from Fig.~\ref{var_r_vs_td} (a). The straight lines are
the best linear fits to the data and in both cases give
$X_L^\text{max}\sim L^{\gamma/\nu}$ with $\gamma/\nu=1.77\pm0.02$.}
\label{FSS_fluct_r}
\end{figure}

In Fig.~\ref{FSS_fluct_r}(a) we plot $\ln(X_L^\text{max})$
versus $\ln(L)$ for $k=0.01$.
A linear fit indicates a power-law divergence of the fluctuations with $L$,
with exponent $\gamma/\nu = 1.77\pm 0.02$.
A different method to extract the power-law exponent, which has some advantage
in eliminating the effects of a nonsingular background term 
(as in $X_L = f + gL^{\gamma/\nu}$ with $f$ and $g$ constants), is to consider 
\begin{equation}
\ln\left[\frac{X_{bL}^\text{max}}{X_L^\text{max}}\right]/\ln b=
\frac{\gamma}{\nu} + {\cal O}(1/\ln b)
\label{eq:bLX}
\end{equation}
with $L$ fixed at a relatively small value (here, $L=80$), and $b>1$. 
For large $L$ and $b$, 
the correction term is proportional to $f/(g \ln b)$, so that the exponent 
can be estimated by plotting the left-hand-side of Eq.~(\ref{eq:bLX}) vs 
$1/\ln b$ and extrapolating to $1/\ln b = 0$, as in 
Fig.~\ref{FSS_fluct_r}(b). The resulting estimate is again 
$\gamma/\nu = 1.77\pm 0.02$.  

\begin{figure}
\centering\includegraphics[clip, scale=0.44]{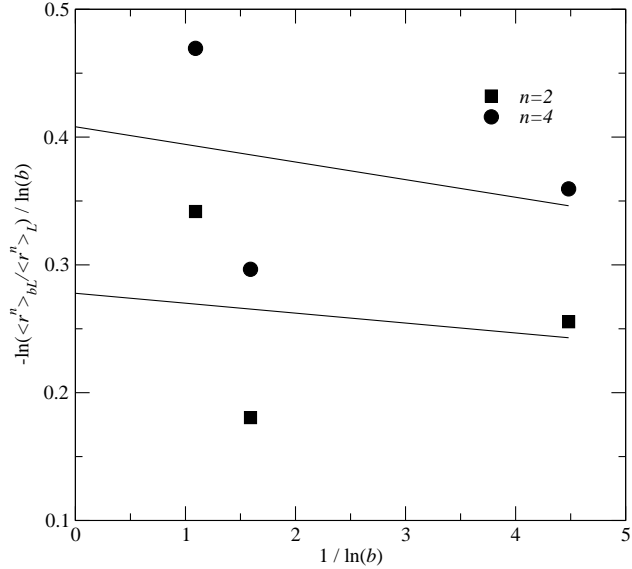}
\caption{Plot of 
$-\ln(\langle r^n \rangle_{bL}/\langle r^n \rangle_L)/\ln(b)$ vs $1/\ln(b)$ 
for $k=0.01$ and
$t_p=230$, including all four system sizes. The straight lines are
the best linear fits to the data, 
giving $\beta/\nu=0.14 \pm 0.06$ for $n=2$
and $\beta/\nu=0.10 \pm 0.03$ for $n=4$.}
\label{r_renorm}
\end{figure}

To obtain an estimate for the exponent ratio $\beta / \nu$, we used the 
scaling relation for the order parameter at the critical point, 
$\langle r^n\rangle\sim L^{- n(\beta/\nu)}$, (again with the possibility of 
a nonscaling background, which can be quite significant for $r$) 
to plot the left-hand side of 
\begin{equation}
-\ln\left[\frac{\langle r^n \rangle_{bL}}{\langle r^n \rangle_L}\right]/\ln b 
= n\frac{\beta}{\nu} + {\cal O}(1/\ln b)
\label{eq:bLr}
\end{equation}
for $L=80$ vs $1/ \ln b$, as shown in Fig.~\ref{r_renorm}.
For lack of a better estimate of the transition point, $\langle r^n \rangle$ 
were measured at the values of $t_d$ corresponding to the maxima of $X_L$.
A linear fit that takes into account all four
system sizes gives $\beta/\nu=0.14\pm 0.06$ for $n=2$ 
and $\beta/\nu=0.10\pm 0.03$ for $n=4$. As a combined estimate, we take 
$\beta/\nu=0.12\pm 0.04$, where the error bar includes some measure of 
our uncertainty about a nonscaling background and other finite-size
effects. 

Combining the exponent estimates we find  
\begin{equation}
2(\beta/\nu)+(\gamma/\nu) = 2.01 \pm 0.03 \approx 2 = d \;,
\end{equation}
where $d$ is the spatial dimension. This result agrees with hyperscaling
\cite{GOLD92}
and indicates that our finite-size scaling results are internally consistent 
\cite{ENDNOTE}. 
Results very similar to the ones presented above were obtained selecting the
period-averaged CO coverage as the order parameter, instead of $r$. 

\begin{figure}
\centering\includegraphics[clip, scale=0.44]{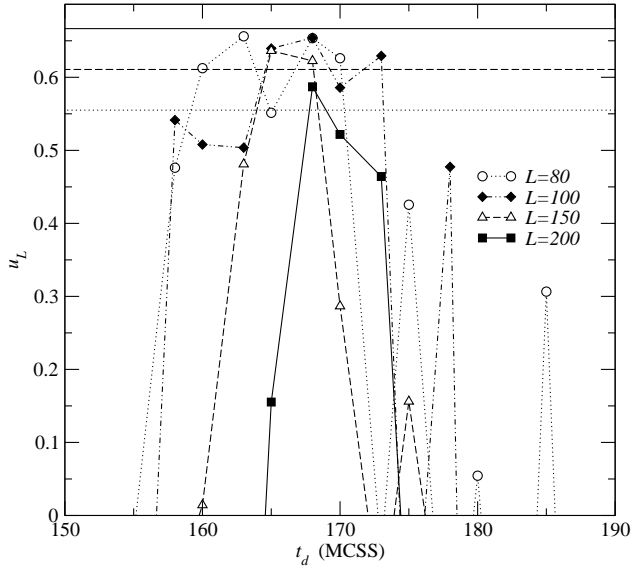}
\caption{
The fourth-order cumulant $u_L$, shown vs $t_d$ for all four system sizes,
$k=0.01$ and $t_p=230$. 
The horizontal lines correspond to $u_L=2/3$ (solid), 
$u_{\rm Ising}^\ast \approx 0.610$ (dashed), and 
$u_{\rm Perc}^\ast \approx 0.555$ (dotted).
See discussion in the text.
The errors are approximated by standard error-propagation methods
and are of the order of 3 \% for the highest values of $u_L$ and of the
order of 14 \% for the lowest.}
\label{fig:cum}
\end{figure}

Our estimated exponent ratios are very close to the analogous two-dimensional 
equilibrium Ising values ($\gamma/\nu = 7/4 = 1.75$ and 
$\beta/\nu = 1/8 = 0.125$ with $\beta = 1$ \cite{GOLD92}), but they are also 
near those for two-dimensional random percolation 
($\gamma/\nu = 43/24 \approx 1.79$ and $\beta/\nu = 5/48 \approx 0.104$ with 
$\nu = 4/3 \approx 1.33$ \cite{STAU92}). 
One way to verify the universality class without calculating $\nu$ directly
(which would require much more accurate data than we have available), is to 
consider another universal quantity, such as the fixed-point value of the 
fourth-order order-parameter cumulant (``Binder cumulant'') \cite{pheno}, 
\begin{equation}
u_L = 1 - 
\frac{\langle (r-\langle r \rangle_L )^4 \rangle_L}
{3 \langle (r - \langle r \rangle_L )^2 \rangle_L^2}
\;,
\label{eq:cum}
\end{equation}
where $\langle \bullet \rangle_L$ denotes the average over the whole time 
series of $r$ for an $L\times L$ system. This cumulant is 
shown vs $t_d$ for different $L$ in Fig.~\ref{fig:cum}. 
The maximum possible value of $u_L$ is 2/3, which is reached in the dynamically 
ordered phase, provided that ergodicity is not broken or 
that $\langle r \rangle_L$ 
is exactly known. This is not so in the present case, and so the cumulant is 
nonmonotonic, becoming negative deep in both the dynamically ordered (large 
$t_d$) and disordered (small $t_d$) phases. 
With sufficiently accurate data, the curves 
representing $u_L$ for different $L$ cross or touch 
at a common point, which represents
an estimate for the critical value of the control variable (here, $t_d$) 
that is quite insensitive to corrections to scaling. The value of $u_L$ 
at this fixed point, $u^\ast$, is a universal quantity characteristic of 
the particular universality class. 
In the present case, the accuracy of our data is not sufficient to use the 
crossing point as an estimate for the critical value of $t_d$ 
(we have instead used the maxima of $X_L$). However, $u^\ast$ is seen 
to be in the vicinity of 0.6, consistent with the very accurately known 
value for the two-dimensional Ising universality class, 
$u_{\rm Ising}^\ast = 0.6106901(5)$ \cite{KAMI93}. 
Even though it is not very accurately determined, the value of $u^\ast$ 
observed for the present 
system is unlikely to be as low as the value for random 
percolation, $u_{\rm Perc}^\ast \approx 0.555$ \cite{ZIFFperc}. 
Our total numerical finite-size scaling evidence thus points 
in the direction that the DPT in this system belongs to the two-dimensional 
equilibrium Ising universality class, together with other DPTs in 
far-from-equilibrium systems, such as the 
one observed in the two-dimensional kinetic Ising model driven by an 
oscillating applied field \cite{ising1,ising2,FUJI01,ENDNOTE2}.

An independent 
indication that the DPT in this system belongs to the equilibrium 
Ising universality class, is a symmetry argument due to 
Grinstein, Jayaprakash, and He \cite{GRIN85}, who argued that the 
equilibrium Ising universality class extends to nonequilibrium cellular 
automata that (i) have local dynamics, (ii) do not conserve the order 
parameter or other auxiliary fields, and (iii) respect the Ising up-down 
symmetry. This result was later extended by Bassler and 
Schmittmann \cite{BASS94}, using renormalization-group arguments, 
to systems that obey conditions (i) and (ii) above, but violate (iii), such 
as some driven lattice gases. The ZGB-k model satisfies these requirements, 
and it should thus belong to the equilibrium Ising class on symmetry grounds. 

\section{Conclusions}

In this paper we have studied 
the dynamic response of the ZGB model with CO desorption 
(the ZGB-k model \cite{tome93}) to
periodic variations of the relative CO pressure $y$, around the
coexistence value $y_2(k)$ that separates the low and high
CO-coverage phases. There is an asymmetry of the lifetimes of the
model: its decontaminating time $\tau_d$ generally differs from the 
contaminating time $\tau_p$. 
We exploited this fact by selecting a square-wave periodic 
CO pressure that stays for a time $t_d$ in the high-production 
region and for a time $t_p$ in the low-production one. We found
that $t_d$ and $t_p$ can be tuned to significantly enhance the
time-averaged catalytic activity of the system beyond its maximum value 
under constant-pressure conditions -- a result we believe should be of 
applied significance.

We also found strong indications that, for sufficiently low values of the
desorption rate, this driven nonequilibrium 
system undergoes a dynamic phase
transition between a dynamic phase of high CO$_2$ production, 
$\langle r \rangle > 0$, and a nonproductive phase 
$\langle r \rangle \approx 0$. 
As the order parameter for this nonequilibrium phase transition we used 
the period-averaged rate of CO$_2$ production, $r$. Our study shows that
the distinction between these phases disappears for a high enough
desorption rate. Applying finite-size scaling techniques in a
similar fashion to what is 
commonly done to study equilibrium second-order phase
transitions, we found that, for small values of the CO desorption rate $k$, 
the fluctuations of the order parameter diverge as a power law 
with the system size, $X_L^\text{max}\sim L^{\gamma/\nu}$
with exponent $\gamma/\nu=1.77\pm0.02$, while moments of 
the order parameter at the transition point decay as 
$\langle r^n \rangle_L \sim L^{- n \beta/\nu}$ with 
$\beta/\nu = 0.12 \pm 0.04$, and the fourth-order 
order-parameter cumulant 
$u_L$ has a fixed-point value of $u^\ast \approx 0.6$. 
These values are close to those of the
two-dimensional Ising universality class, and together with general 
symmetry arguments \cite{GRIN85,BASS94} they represent reasonable evidence that 
this far-from equilibrium phase transition belongs to the same universality 
class as the equilibrium Ising model. 
A a higher level of confidence about the universality class 
would require simulations at least an order of magnitude more 
extensive than the ones presented here -- a task beyond the scope of the 
present paper. 

Finally, we note that the
enhancement of the CO$_2$ production and the continuous DPT 
are probably closely related phenomena 
since the critical cluster associated with the phase transition is likely to 
provide more empty sites available for O$_2$ adsorption near adsorbed CO 
molecules, than the sharp interfaces expected near the first-order 
coexistence line seen under constant-$y$ conditions.

\begin{acknowledgments}
This work was supported in part by U.S.\ National Science 
Foundation Grant No.\ DMR-0240078 at Florida State University and 
Grant No.\ DMS-0244419 at The University of Michigan. 
\end{acknowledgments}


\begin{thebibliography}{99}

\bibitem{general1} H.~J.\ Jensen, \textit{Self-Organized
Criticality: Emergent Complex Behavior in Physical and Biological
Systems} (Cambridge University Press, Cambridge, England, 1998).

\bibitem{general2} J.~Marro and R.~Dickman,
\textit{Nonequilibrium Phase Transitions in Lattice Models}
(Cambridge University Press, Cambridge, England, 1999).

\bibitem{surface} K.~Christmann, \textit{Introduction to Surface
Physical Chemistry} (Steinkopff Verlag, Darmstadt 1991); V.~P.~Z.\
Zhdanov and B.~Kazemo, Surf.\ Sci.\ Rep.\ {\bf 20}, 111 (1994).

\bibitem{catalytic} G.~C.\ Bond, \textit{Catalysis: Principles and
Applications} (Clarendon, Oxford, 1987).

\bibitem{imbhil95} R.~Imbhil and G.~Ertl, Chem.\ Rev.\
\textbf{95}, 697 (1995).

\bibitem{ehsasi89} M.~Ehsasi, M.~Matloch, O.~Frank, J.~H.\ Block, 
K.~Christmann, F.~S.\ Rys and W.~Hirschwald, J.\ Chem.\ Phys.\
\textbf{91}, 4949 (1989).

\bibitem{vaporciyan88} G.~Vaporciyan, A.~Annapragada, and E.~Gulari, 
Chem.\ Eng.\ Sci.\ \textbf{43}, 2957 (1988).

\bibitem{cutlip83} M.~B.\ Cutlip, C.~J.\ Hawkins, D.~Mukesh, W.~Morton, 
and C.~N.\ Kenney,
Chem.\ Eng.\ Commun.\ \textbf{22}, 329 (1983).

\bibitem{hegedus80} L.~L.\ Hegedus, C.~C.\ Chang, D.~J.\ McEwen, 
and E.~M.\ Sloan,
Ind.\ Eng.\ Chem.\ Fundam.\ \textbf{19}, 367 (1980).

\bibitem{matsushima79} T.~Matsushima, H.~Hashimoto and I. Toyoshima, J.\ Catal.\ \textbf{58},
303 (1979).

\bibitem{golchet78} A.~Golchet and J.~M.~White, J.\ Catal.\ \textbf{53},
266 (1978).

\bibitem{christmann73} K.~Christmann and G. Ertl, Z.\ Naturforsch.\ Teil\ A
\textbf{28}, 1144 (1973).

\bibitem{ziff86} R.~M.\ Ziff, E.~Gulari, and Y.~Barshad, 
Phys.\ Rev.\ Lett.\ \textbf{56}, 2553 (1986).

\bibitem{lopez00} A.~C.\ L\'opez and E.~V.\ Albano, J.\ Chem.\ Phys.\
\textbf{112}, 3890 (2000).

\bibitem{TAMM98}
M.~Tammaro and J.~W.\ Evans, J.\ Chem.\ Phys.\ \textbf{108}, 762 (1998). 

\bibitem{tome93} T.~Tom\'e and R.~Dickman, Phys.\ Rev.\ E
\textbf{47}, 948 (1993).

\bibitem{machado04} E.~Machado, G.~M.\ Buend\y a, and P.~A.\ Rikvold, 
\textit{in preparation}.

\bibitem{TOME90}
T.~Tom{\'e} and M.~J.\ de~Oliveira, Phys.\ Rev.\ A \textbf{41}, 4251 (1990). 

\bibitem{MEND91}
J.~F.~F.\ Mendes and J.~S.\ Lage, J.\ Stat.\ Phys.\ \textbf{64}, 653 (1991). 

\bibitem{ACHAR97_1}
M.~Acharyya, Phys.\ Rev.\ E\ \textbf{56}, 1234 (1997).

\bibitem{ACHAR97_2}
M.~Acharyya, Phys.\ Rev.\ E\ \textbf{56}, 2407 (1997).

\bibitem{CHAK99}
B.~Chakrabarti and M.~Acharyya, Rev.\ Mod.\ Phys.\ \textbf{71}, 847 (1999). 

\bibitem{ising1}
S.~W.\ Sides, P.~A.\ Rikvold and M.~A.\ Novotny,
Phys.\ Rev.\ Lett.\ \textbf{81}, 834 (1998);
Phys.\ Rev.\ E \textbf{59}, 2710 (1999).

\bibitem{ising2} G.~Korniss, C.~J.\ White, P.~A.\ Rikvold and M.~A.\ Novotny,
Phys.\ Rev.\ E \textbf{63}, 016120 (2000).

\bibitem{FUJI01}
H.~Fujisaka, H.~Tutu, and P.~A.\ Rikvold, Phys.\ Rev. E \textbf{63}, 
016120 (2001); Erratum: \textbf{63} 059903(E) (2001). 

\bibitem{KORN02}
G.~Korniss, P.~A.\ Rikvold, and M.~A.\ Novotny, Phys.\ Rev. E \textbf{66},
056127 (2002). 

\bibitem{JANG01}
H.~Jang and J.~Grimson, Phys.\ Rev. E \textbf{63}, 066119 (2001).

\bibitem{JANG03}
H.~Jang, J.~Grimson, and C.~K.\ Hall, Phys.\ Rev.\ B \textbf{67}, 094411 (2003);
{\bf 68}, 046115 (2003). 

\bibitem{YASU02}
T.~Yasui, H.~Tutu, M.~Yamamoto, and H.~Fujisaka, Phys.\ Rev.\ E \textbf{66}, 
036123 (2002); Erratum: \textbf{67}, 019901(E) (2003). 

\bibitem{pheno} K.~Binder, in \textit{Finite Size Scaling 
and Numerical Simulation of Statistical
Systems}, edited by V.~Privman (World Scientific, Singapore, 1990).

\bibitem{GOLD92}
See, e.g., 
N.~Goldenfeld, \textit{Lectures on Phase Transitions and the Renormalization 
Group\/} (Addison-Wesley, Reading, MA, 1992). 

\bibitem{ENDNOTE}
We also tried to perform fits to determine the exponent ratios, excluding 
the results for the smallest system size, $L=80$. We did this by using 
Eqs.~(\ref{eq:bLX}) and~(\ref{eq:bLr}) with $L=100$. The resulting exponent 
estimates, $\gamma/\nu \approx 1.69$, $\beta/\nu \approx 0.35$ for $n=2$, 
and $\beta/\nu \approx 0.21$ for $n=4$, are strongly $n$-dependent and 
significantly violate the hyperscaling relation: 
$2(\beta/\nu)+\gamma/\nu \approx 2.4$ for $n=2$ and 
$2(\beta/\nu)+\gamma/\nu \approx 2.1$ for $n=4$.
These fits are thus not internally consistent, probably because the 
uncertainty in our data is such that the maximum range of system sizes is 
required to obtain reasonably accurate exponent ratios. 

\bibitem{STAU92}
D.~Stauffer and A.~Aharony, \textit{Introduction to Percolation Theory\/}, 
2nd Ed.\ (Taylor \& Francis, London, 1992). 

\bibitem{KAMI93}
G.~Kamieniarz and H.~W.~J.\ Bl{\"o}te, 
J.\ Phys.\ A: Math.\ Gen.\ \textbf{26}, 201 (1993). 

\bibitem{ZIFFperc}
We were unable to find the fixed-point value of the fourth-order cumulant 
for random percolation in the literature, and we therefore calculated it 
ourselves by Monte Carlo simulation. Specifically, 
we used bond percolation on an $L \times L$ square lattice with 
a square boundary and used as the order parameter the probability that a 
randomly chosen bond belongs to the percolating cluster. For $L=100$ and 
200, the cumulants touch at the exactly known critical 
$p_c = 1/2$, giving $u_{\rm Perc}^\ast = 0.5555 \pm 0.0005$. 

\bibitem{ENDNOTE2}
The similarity with the Ising model with its ``up-down symmetry'' is most easily
seen by defining the effective ``magnetization'' variable $S=2\theta_\text{CO}-1$
as in the standard Ising/lattice gas mapping.

\bibitem{GRIN85}
G.~Grinstein, C.~Jayaprakash, and Y.~ He, Phys.\ Rev.\ Lett.\ \textbf{55}, 
2527 (1985). 

\bibitem{BASS94}
K.~E.\ Bassler and B.~Schmittmann, Phys.\ Rev.\ Lett.\ \textbf{73}, 3343 (1994). 


\end{thebibliography}
\end{document}